\documentclass[aps,prl,amsmath,amssymb,footinbib,showpacs,twocolumn]{revtex4-2}
\usepackage{amsmath}
\usepackage{amssymb}
\usepackage{amsthm}
\usepackage{setspace}
\usepackage{graphicx}
\usepackage{braket}
\usepackage{mathrsfs}
\usepackage{float}
\usepackage[colorlinks = true,linkcolor = blue,urlcolor  = blue,citecolor = blue,anchorcolor = blue]{hyperref}
\usepackage[utf8]{inputenc}
\usepackage[english]{babel}
\usepackage{bm}
\usepackage{csquotes}
\usepackage{hyperref}

\begin{document}

\title{Chiral anomaly and longitudinal magnetoconductance in pseudospin-1 fermions}
\author{Azaz Ahmad}
\affiliation{School of Physical Sciences, Indian Institute of Technology Mandi, Mandi 175005, India.}
\author{Gargee Sharma}
\affiliation{School of Physical Sciences, Indian Institute of Technology Mandi, Mandi 175005, India.} 

\begin{abstract}
Chiral anomaly (CA), a hallmark of Weyl fermions, has emerged as a cornerstone of condensed matter physics following the discovery of Weyl semimetals. While the anomaly in pseudospin-1/2 (Weyl) systems is well-established, its extension to higher-pseudospin fermions remains a frontier with critical implications for transport phenomena in materials with multifold fermions.  We present a rigorous quasiclassical analysis of CA and longitudinal magnetotransport in pseudospin-1 fermions, advancing beyond conventional models that assume constant relaxation times and neglect the orbital magnetic moment and global charge conservation. Our study uncovers a magnetic-field dependence of the longitudinal magnetoconductance: it is positive and quadratic-in-$B$ for weak internode scattering and transitions to negative values beyond a critical internode scattering strength. Notably, the critical threshold is lower for pseudospin-1 fermions compared to their pseudospin-1/2 counterparts. We show analytically that the zero-field conductivity is affected more strongly by internode scattering for pseudospin-1 fermions than conventional Weyl fermions. 
These insights provide a foundational framework for interpreting recent experiments on multifold fermions and offer a roadmap for probing CA in candidate materials with space group symmetries 
199, 
214, and 
220. 
\end{abstract}

\maketitle

\textit{Introduction:} Chiral anomaly (CA) of Weyl fermions was first discovered as a contributing factor in the decay of pions~\cite{bertlmann2000anomalies}. Over the past decade, chiral anomaly has seen a remarkable resurgence in condensed matter physics, driven by the definitive discovery of Weyl fermions in solids~\cite{hosur2013recent,armitage2018weyl,Yan_2017,hasan2017discovery,burkov2018weyl,ong2021experimental,nagaosa2020transport,lv2021experimental,mandal2022chiral,ahmad2024geometry}. Theoretical advances, however, date back to the 1980s when Nielsen \& Ninomiya, who first studied lattice Weyl fermions, proved that they must occur in pairs of opposite chiralities~\cite{nielsen1981no,nielsen1983adler}. Such pairing ensures the conservation of both the chiral and global charge. When external gauge fields are present, chiral charge is not conserved, a phenomenon that is now well known as the chiral anomaly or the Adler-Bell-Jackiw (ABJ) anomaly~\cite{adler1969axial} of Weyl fermions. The manifestation of this anomaly is investigated through transport, thermoelectric, and optical experiments in systems hosting Weyl fermions, known as Weyl semimetals (WSMs)~\cite{parameswaran2014probing,hosur2015tunable,goswami2015optical,goswami2013axionic,son2013chiral,burkov2011weyl,burkov2014anomalous,lundgren2014thermoelectric,sharma2016nernst,kim2014boltzmann,zyuzin2017magnetotransport,cortijo2016linear,das2019berry,kundu2020magnetotransport}.

While Weyl fermions have made an entry from high-energy physics to condensed matter, certain symmetries of condensed matter systems make it possible to realize free fermionic excitations which are not allowed by Poincar\'{e} symmetry in high-energy physics~\cite{bradlyn2016beyond,tang2017multiple,chang2018topological}. Specifically, the low-energy Hamiltonian of a Weyl fermion is $H_\mathbf{k}\sim\chi\mathbf{k}\cdot\boldsymbol{\sigma}$, where $\boldsymbol{\sigma}$ is the vector of Pauli matrices, and $\chi$ is the chirality of the fermion. Fermionic excitations of the type $H_\mathbf{k}\sim \chi\mathbf{k}\cdot\mathbf{S}$ are allowed in solids, where $\mathbf{S}$ is a higher (pseudo)spin generalization of the vector of Pauli matrices. These excitations are multifold degenerate chiral quasiparticles, carry a nontrivial Chern number $|\mathcal{C}|>1$, and are sources and sinks of the Berry curvature. In contrast, Weyl fermions have a twofold degeneracy with Chern number $|\mathcal{C}|=1$.

Although CA has been heavily discussed in the context of Weyl fermions or even Kramer-Weyl fermions~\cite{cheon2022chiral,das2023chiral,varma2024magnetotransport}, its generalization beyond fermions of pseudospin$-1/2$ has received little attention. On generic grounds, chiral particles respond differently to external magnetic fields, and
one does expect the anomaly to persist. When external fields are applied such that $\mathbf{E}\cdot\mathbf{B}\neq 0$, the chiral charge is not conserved, and a chiral current is generated. There have been theoretical advances based on this idea, aimed at understanding chiral anomaly and associated transport features in multifold fermions~\cite{ezawa2017chiral,lepori2018axial,nandy2019generalized}. Still, the quasiclassical analysis suffers from shortcomings such as imposing a constant relaxation time, neglecting orbital magnetic moment, internode scattering effects, and global charge conservation. Going beyond these standard assumptions is indispensable to correctly reproduce the physics of chiral anomaly and magnetotransport in chiral fermions in the experimentally accessible low-field limit as shown in some recent works~\cite{knoll2020negative,sharma2020sign,ahmad2021longitudinal,ahmad2023longitudinal}.

This Letter presents a complete quasiclassical analysis of chiral anomaly and longitudinal magnetotransport in pseudospin-1 fermions. Moving beyond the conventional constant relaxation-time approximation and taking into account the effects of orbital magnetic moment and global charge conservation, we discover that chiral anomaly in pseudospin-1  fermions generates a positive and quadratic-in-$B$ longitudinal magnetoconductance for low internode scattering, which becomes negative as the strength of internode scattering ($\alpha$) is increased beyond a critical value ($\alpha^{(1)}$). Interestingly, this critical intervalley scattering strength is lower in pseudospin-1 fermions than the regular Weyl-fermions ($\alpha^{(1)}<\alpha^{(1/2)}$). We also show analytically that the zero-field conductivity decreases sharply with internode scattering for pseudospin-1 fermions than conventional Weyl fermions.
This study becomes even more pertinent in light of recent experiments that have probed this anomaly in multifold fermionic systems~\cite{balduini2024intrinsic}, and upcoming experiments that could be performed on relevant materials belonging to space groups $199$, $214$, and $220$~\cite{bradlyn2016beyond}.  

\textit{Semimetals with 3-fold degeneracy:} The low-energy Hamiltonian of the pseudospin-1 fermion expanded about a nodal point is~\cite{bradlyn2016beyond}:
\begin{equation}
H(\mathbf{k})=\hbar v_{F}\mathbf{k}\cdot\mathbf{S}.
\label{Hamiltonian}    
\end{equation}
where $v_{F}$ is a material-dependent velocity parameter, $\mathbf{k}$ is the wave vector measured from the nodal point, and $\mathbf{S}$ is the vector of spin-1 Pauli matrices. One may diagonalize this Hamiltonian to get energy eigenvalues:
\begin{align}
\epsilon_k = 0, ~\pm{\hbar v_{F}}k.
\label{Dispersion}
\end{align}
The dispersion consists of three bands, including a flat band with zero energy.
The Bloch-states corresponding to the non-zero energies are calculated to be:
\begin{align}
|u^+\rangle = \left[\cos^2(\theta) e^{-2i\phi}, ~ \frac{1}{\sqrt{2}}\sin(\theta) e^{-i\phi},  ~\sin^2(\theta/2)\right]^\mathrm{T}, \nonumber\\
|u^-\rangle = \left[\sin^2(\theta/2) e^{-2i\phi}, ~ \frac{-1}{\sqrt{2}}\sin(\theta) e^{-i\phi},  ~\cos^2(\theta/2)\right]^\mathrm{T},
\label{Eq:wave_function}
\end{align}
with $\theta$ and $\phi$ being the polar and azimuthal angles respectively. The Chern numbers of the bands are $\nu=\mp 2$, while the flat band is trivial. 
Such fermions may emerge in a body-centered cubic lattice with space groups 199, 214, and 220 at the $\mathbf{P}$ point of the BZ~\cite{bradlyn2016beyond}. The $\mathbf{P}$ point does not map to its time-reversal partner ($\mathbf{P}\neq -\mathbf{P}$).
Due to the Nielsen-Ninomiya theorem, the Brillouin zone must compensate for sources and sinks of the Berry curvature~\cite{nielsen1981no,nielsen1983adler}. Therefore, the low-energy minimal Hamiltonian for pseudospin-1 semimetal may be written as 
\begin{align}
    H(\mathbf{k})=\sum_{\chi=\pm 1} \chi \hbar v_{F}\mathbf{k}\cdot\mathbf{S}.
    \label{Eq:Hamiltonian}
\end{align}
The Berry curvature of the conduction bands of the Hamiltonian is evaluated to be:
\begin{align}
  \boldsymbol{\Omega}^\chi_\mathbf{k} = i \nabla_{\mathbf{k}} \times \langle u^{\chi}(\mathbf{k}) | \nabla_{\mathbf{k}} | u^{\chi}(\mathbf{k}) \rangle \equiv -\chi \mathbf{k} /k^3.
    \label{Eq:BC}
\end{align}
In contrast to a classical point particle, a Bloch wave packet in a crystal possesses a finite spatial extent. Consequently, it undergoes self-rotation around its center of mass, resulting in an orbital magnetic moment (OMM), expressed as ~\cite{xiao2010berry,hagedorn1980semiclassical,chang1996berry},
\begin{align}
\mathbf{m}^{\chi}_\mathbf{k}= -\frac{ie}{2\hbar} \text{Im} \langle \nabla_{\mathbf{k}} u^{\chi}|[ \epsilon_0(\mathbf{k}) - \hat{H}^{\chi}(\mathbf{k}) ]| \nabla_{\mathbf{k}} u^{\chi}\rangle \nonumber\\
=-{\chi e v_{F} \mathbf{k}}/{k^2}.
\label{Eq:OMM}
\end{align}
In the presence of an external magnetic field ($\mathbf{B}$), the orbital magnetic moment couples to it, and the dispersion relation is modified as $\epsilon^\chi_\mathbf{k} \rightarrow \epsilon^\chi_\mathbf{k} - \mathbf{m}^{\chi}_\mathbf{k} \cdot \mathbf{B}$. Consequently, the Fermi contour becomes anisotropic: 
\begin{align} k_F^{\chi}(\theta) = \frac{\epsilon_{F} + \sqrt{\epsilon^{2}_{F} - \chi \eta ~\epsilon^2_0 \cos(\theta)}}{2 v_{F}\hbar}.
\label{Eq:K_chi}
\end{align}
with $\epsilon_F$ being the Fermi energy, $\epsilon_0 = \sqrt{4 eB  v^2_{F}\hbar}$ and the variable $\eta\in\{0, 1\}$ is used to toggle the effect of the orbital magnetic moment, such that its effect can be studied separately. 

\textit{Quasiclassical transport:} The dynamics of the quasiparticles in the presence of electric ($\mathbf{E}$) and magnetic ($\mathbf{B}$) fields, are described by the following equation~\cite{son2012berry,knoll2020negative}:
\begin{align}
\dot{\mathbf{r}}^\chi &= \mathcal{D}^\chi_\mathbf{k} \left( \frac{e}{\hbar}(\mathbf{E}\times \mathbf{\Omega}^\chi) + \frac{e}{\hbar}(\mathbf{v}^\chi\cdot \boldsymbol{\Omega}^\chi) \mathbf{B} + \mathbf{v}_\mathbf{k}^\chi\right) \nonumber\\
\dot{\mathbf{p}}^\chi &= -e \mathcal{D}^\chi_\mathbf{k} \left( \mathbf{E} + \mathbf{v}_\mathbf{k}^\chi \times \mathbf{B} + \frac{e}{\hbar} (\mathbf{E}\cdot\mathbf{B}) \boldsymbol{\Omega}^\chi \right).
\label{Couplled_equation}
\end{align}
To describe the dynamics of three-dimensional pseudospin-1 fermions under external electric and magnetic fields, we employ the quasiclassical Boltzmann formalism, where the evolution of the non-equilibrium distribution function $f^{\chi}_{\mathbf{k}}$ is given by:
\begin{align}
\dfrac{\partial f^{\chi}_{\mathbf{k}}}{\partial t}+ {\Dot{\mathbf{r}}^{\chi}_{\mathbf{k}}}\cdot \mathbf{\nabla_r}{f^{\chi}_{\mathbf{k}}}+\Dot{\mathbf{k}}^{\chi}\cdot \mathbf{\nabla_k}{f^{\chi}_{\mathbf{k}}}=I_{\mathrm{coll}}[f^{\chi}_{\mathbf{k}}],
\label{MB_equation}
\end{align}
with $f^{\chi}_\mathbf{k} = f_{0} + g^{\chi}_{\mathbf{k}}$, where $f_{0}$ is the Fermi-Dirac distribution, and $g^{\chi}_{\mathbf{k}}$ is the deviation. We fix the electric and magnetic fields along the $z-$direction and the deviation up to the first order in perturbation is expressed as:
\begin{align}
g^{\chi}_\mathbf{k}&= -e\left({\dfrac{\partial f_{0}}{\partial {\epsilon}}}\right){\Lambda^{\chi}_\mathbf{k}} E,
\label{Eq:g1}
\end{align}
where ${\Lambda}^{\chi}_\mathbf{k}$ is unknown function to be evaluated. The collision integral considers two impurity-dominated distinct scattering processes: (i) scattering between $\chi$ and $\chi'\neq\chi$, and (ii) scattering between $\chi$ to $\chi'=\chi$. These are also known as internode (intervalley) scattering, and intranode (intravalley) scattering, respectively. The collision integral is expressed as~\cite{son2013chiral,kim2014boltzmann}:
\begin{align}
 I_{coll}[f^{\chi}_{\mathbf{k}}]=\sum_{\chi' \mathbf{k}'}{\mathbf{W}^{\chi \chi'}_{\mathbf{k k'}}}{(f^{\chi'}_{\mathbf{k'}}-f^{\chi}_{\mathbf{k}})},
 \label{Collision_integral}
\end{align}
where the scattering rate ${\mathbf{W}^{\chi \chi'}_{\mathbf{k k'}}}$ calculated using Fermi's golden rule:
\begin{align}
\mathbf{W}^{\chi \chi'}_{\mathbf{k k'}} = \frac{2\pi n}{\mathcal{V}}|\bra{u^{\chi'}(\mathbf{k'})}U^{\chi \chi'}_{\mathbf{k k'}}\ket{u^{\chi}(\mathbf{k})}|^2\times\delta(\epsilon^{\chi'}(\mathbf{k'})-\epsilon_F).\nonumber\\
\label{Fermi_gilden_rule}
\end{align}
Here $n$ represents the impurity concentration, $\mathcal{V}$ represents the system volume, and $U^{\chi \chi'}_{\mathbf{k k'}}$ describes the scattering potential profile. For elastic impurities, we set $U^{\chi \chi'}_{\mathbf{k k'}}= I_{3\times3}U^{\chi \chi'}$, where $U^{\chi \chi'}$ controls scattering between electrons of different ($\chi\neq\chi'$) and same chirality ($\chi=\chi'$). The relative magnitude of chirality-breaking to chirality-preserving scattering is expressed by the ratio $\alpha = U^{\chi\chi'\neq\chi}/U^{\chi\chi}$. The overlap $\mathcal{T}^{\chi \chi'}({\theta,\theta',\phi,\phi'})=|\bra{u^{\chi'}(\mathbf{k'})}U^{\chi \chi'}_{\mathbf{k k'}}\ket{u^{\chi}(\mathbf{k})}|^2$ is generally a function of both the polar and azimuthal angles, making the scattering strongly anisotropic. However, for the chosen arrangement of the fields, i.e., $\mathbf{E}=E\hat{z}$, and  $\mathbf{B} = B \hat{z} $, the terms involving $\phi$, and $\phi'$ are irrelevant due to azimuthal symmetry as they vanish when the integral with respect to $\phi'$ is performed. We drop such terms and write the the overlap function as 
\begin{align}
\mathcal{T}^{\chi \chi'}_{\theta,\theta}
=\begin{cases}
			[\cos^4(\theta/2) \cos^4(\theta'/2) + \sin^4(\theta/2) \sin^4(\theta'/2)\\+ \frac{1}{4} \sin^2(\theta) \sin^2(\theta')]
            \delta_{\chi,\chi'}+\\
            [\sin^4(\theta/2) \cos^4(\theta'/2) + \cos^4(\theta/2) \sin^4(\theta'/2)\\+ \frac{1}{4} \sin^2(\theta) \sin^2(\theta')]\delta_{\chi,-\chi'}
		 \end{cases}
   \label{Overlap_of_spinor}
\end{align}

Using Eq's.~\ref{Couplled_equation}, \ref{Eq:g1} and \ref{Collision_integral}, Eq.~\ref{MB_equation} is written in the following form:
\begin{align}
&\mathcal{D}^{\chi}_\mathbf{k}\left[{v^{\chi,z}_{\mathbf{k}}}+\frac{eB}{\hbar}(\mathbf{v^{\chi}_k}\cdot\mathbf{\Omega}^{\chi}_k)\right]
 = \sum_{\chi' \mathbf{k}'}{\mathbf{W}^{\chi \chi'}_{\mathbf{k k'}}}{(\Lambda^{\chi'}_{\mathbf{k'}}-\Lambda^{\chi}_{\mathbf{k}})}.
 \label{Eq_boltz_E1}
 \end{align}
 Before further simplifying the above equation, we define the chiral scattering rate as follows:
\begin{align}
\frac{1}{\tau^{\chi}(\theta)}=\sum_{\chi'}\mathcal{V}\int\frac{d^3\mathbf{k'}}{(2\pi)^3}(\mathcal{D}^{\chi'}_{\mathbf{k}'})^{-1}\mathbf{W}^{\chi \chi'}_{\mathbf{k k'}}.
\label{Tau_invers}
\end{align}
Eq.~\ref{Eq_boltz_E1} then transforms to:
\begin{align}
h^{\chi}(\theta) + \frac{\Lambda^{\chi}(\theta)}{\tau^{\chi}(\theta)}=\sum_{\chi'}\mathcal{V}\int\frac{d^3\mathbf{k}'}{(2\pi)^3} \mathcal{D}^{\chi'}_{\mathbf{k}'}\mathbf{W}^{\chi \chi'}_{\mathbf{k k'}}\Lambda^{\chi'}(\theta').
\label{MB_in_term_Wkk'}
\end{align}
Here, $h^{\chi}(\theta)=\mathcal{D}^{\chi}_{\mathbf{k}}[v^{\chi}_{z,\mathbf{k}}+eB(\mathbf{\Omega}^{\chi}_{k}\cdot \mathbf{v}^{\chi}_{\mathbf{k}})/
\hbar]$, evaluated on the Fermi surface. Employing azimuthal symmetry, Eq.~\ref{Tau_invers} and Eq.~\ref{MB_in_term_Wkk'} simplify to an integration over $\theta'$:
\begin{align}
\frac{1}{\tau^{\chi}(\theta)} =  \mathcal{V}\sum_{\chi'} \Pi^{\chi\chi'}\int\frac{(k')^3\sin{\theta' d\theta'}}{|\mathbf{v}^{\chi'}_{k'}\cdot{\mathbf{k'}^{\chi'}}|} \mathcal{T}^{\chi\chi'}_{\theta \theta'}(\mathcal{D}^{\chi'}_{\mathbf{k'}})^{-1}.
\label{Tau_inv_int_theta}
\end{align}
\begin{align}
&h^{\chi}(\theta) + \frac{\Lambda^{\chi}(\theta)}{\tau^{\chi}(\theta)}\nonumber\\&=\mathcal{V}\sum_{\chi'} \Pi^{\chi\chi'}\int d\theta' f^{\chi'}(\theta') \mathcal{T}^{\chi\chi'}_{\theta \theta'} \Lambda^{\chi'}(\theta')/\tau^{\chi'}(\theta'),
\label{Eq_h_tau_boltz}
\end{align}
 where, $\Pi^{\chi \chi'} = N|U^{\chi\chi'}|^2 / 4\pi^2 \hbar^2$, $f^{\chi} (\theta)=\frac{(k)^3}{|\mathbf{v}^\chi_{\mathbf{k}}\cdot \mathbf{k}^{\chi}|} \sin\theta (\mathcal{D}^\eta_{\mathbf{k}})^{-1} \tau^\chi(\theta)$ and $\mathcal{T}^{\chi\chi'}_{\theta \theta'}$ is defined in Eq.~\ref{Overlap_of_spinor}. With the ansatz $\Lambda^{\chi}(\theta)= [\lambda^{\chi}-h^{\chi}(\theta) + a^{\chi}\cos^4(\theta/2) +b^{\chi}\sin^4(\theta/2)+c^{\chi}\sin^2(\theta)]\tau^{\chi}(\theta)$, Eq.~\ref{Eq_h_tau_boltz} is expressed as:
 \begin{multline}
\lambda^{\chi} + a^{\chi}\cos^4(\theta/2) +b^{\chi}\sin^4(\theta/2)+c^{\chi}\sin^2(\theta)\\
=\mathcal{V}\sum_{\chi'}\Pi^{\chi\chi'}\int f^{\chi'}(\theta')d\theta' ~\mathcal{T}^{\chi\chi'}_{\theta \theta'} \\\times[\lambda^{\chi'}-h^{\chi'}(\theta') + a^{\chi'}\cos^4(\theta'/2) +b^{\chi'}\sin^4(\theta'/2)+c^{\chi'}\sin^2(\theta')].\\
\label{Boltzman_final}
\end{multline}
When explicitly written, this equation consists of eight simultaneous equations that need to be solved for eight variables, and particle number conservation serves as an additional constraint.  
The current is calculated using the expression :
\begin{align}
    \mathbf{J}=-e\sum_{\chi,\mathbf{k}} f^{\chi}_{\mathbf{k}} \dot{\mathbf{r}}^{\chi}.
    \label{Eq:J_formula}
\end{align}
\begin{figure}
    \centering
    \includegraphics[width=\columnwidth]{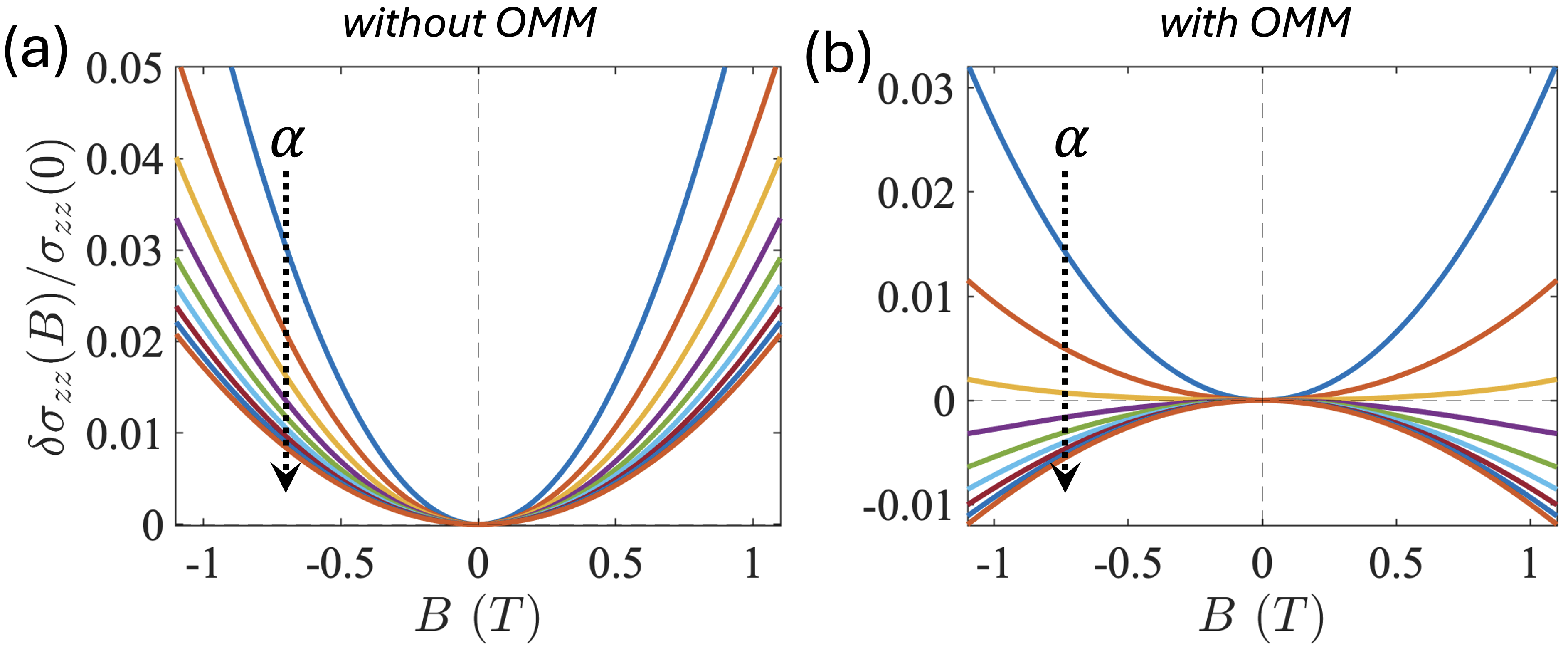}
    \caption{Longitudinal magnetoconductivity in a pseudospin-1 semimetal with and without including the effect of orbital magnetic moment. Increasing the relative magnitude of the internode scattering strength ($\alpha$) results in a reversal of the sign of LMC from positive to negative.}
    \label{fig:lmc:spin1}
\end{figure}
\textit{Results and Discussion:} We study the longitudinal magnetoconductance of the pseudospin-1 semimetal subjected to parallel electric and magnetic fields ($\mathbf{E}\parallel\mathbf{B}$) in the semiclassical regime. This specific field arrangement allows us to restrict the current flow along the $z$-direction, making it easy to make analytical progress. For parallel electric and magnetic fields, the Lorentz force ($\mathbf{F}\propto \mathbf{v}\times\mathbf{B}$) is zero, which makes no conventional transverse current possible in this orientation. LMC is defined as
\begin{align}
\delta\sigma_{zz}(B) = \sigma_{zz} (B) - \sigma_{zz} (0). 
\end{align} 
In Fig.~\ref{fig:lmc:spin1} we plot LMC (normalized with respect to the zero-field conductivity) as a function of the magnetic field. The conductivity is quadratic in the magnetic field. When $\alpha<\alpha_c^{(1)}$, the conductivity is positive, and switches sign when $\alpha>\alpha_c^{(1)}$, where $\alpha_c^{(1)}$ is the critical relative internode scattering strength. When the effects of orbital magnetic moment are turned off, we only obtain positive LMC irrespective of $\alpha$. Increasing $\alpha$ only reduces the magnitude of the conductivity but does not change the sign. Before we discuss further, it is useful to compare and contrast our results with a conventional spin-$1/2$ Weyl semimetal.   
Son \& Spivak~\cite{son2013chiral} first predicted positive and quadratic LMC in Weyl semimetals driven by chirality-breaking internode scattering. However, several recent studies later proposed that intranode scattering, by itself, could lead to positive longitudinal magnetoconductivity by including a force term $\mathbf{E}\cdot\mathbf{B}$ in equations of motion~\cite{kim2014boltzmann,lundgren2014thermoelectric,cortijo2016linear,sharma2016nernst,zyuzin2017magnetotransport,das2019berry,kundu2020magnetotransport}. However, these results do not account for global charge conservation, orbital magnetic moment, and assume a constant relaxation time, both of which are crucial to obtain the correct magnetoconductivity results. Going beyond the standard approximations reveals that LMC in Weyl fermions is negative for zero internode scattering~\cite{sharma2023decoupling}, positive for infinitesimal internode scattering, and negative if the internode scattering is large enough~\cite{knoll2020negative,sharma2023decoupling}. The results obtained here for pseudospin-1 fermions also paint a similar picture, but a closer comparison sheds further light on their universal features.
\begin{figure}
    \centering
    \includegraphics[width=\columnwidth]{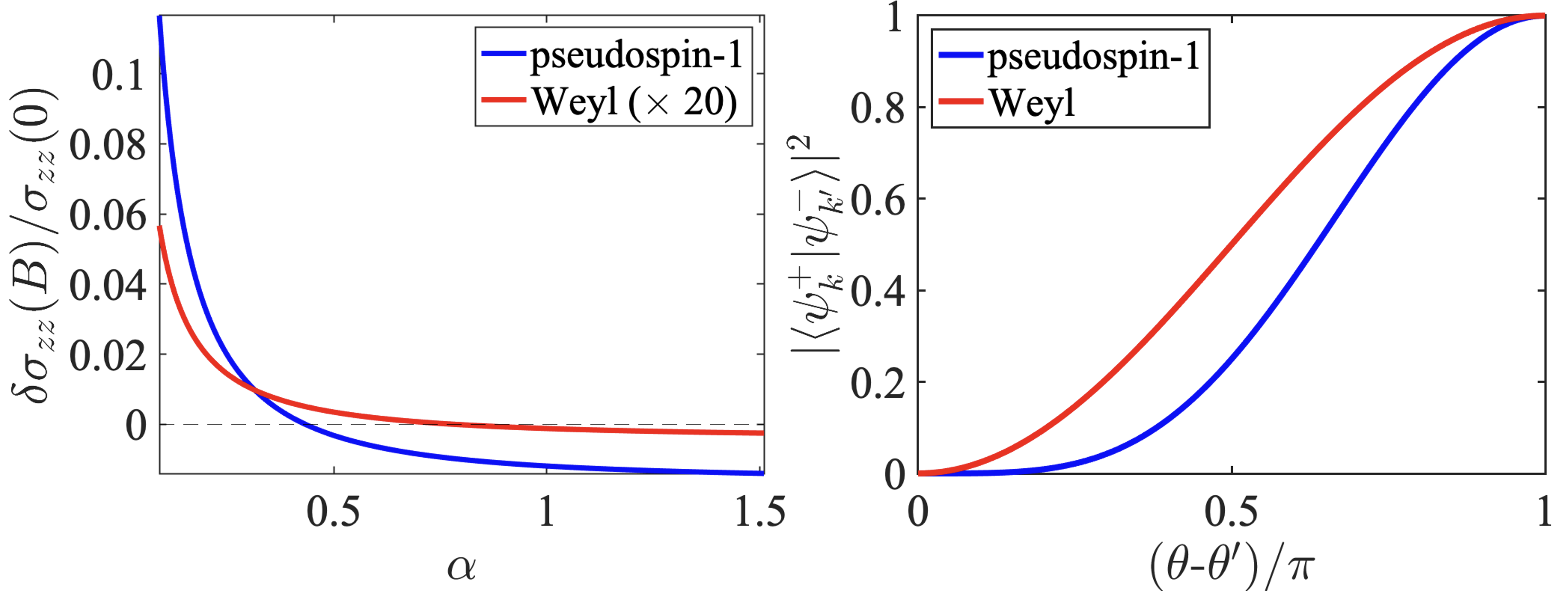}
    \caption{(a) Longitudinal magnetoconductance for a fixed magnetic field as a function of the intervalley scattering strength $\alpha$. The switch from positive to negative LMC happens at a higher value for Weyl fermions compared to pseudospin-1. (b) Overlap between fermions at different valleys $|\langle \psi_\mathbf{k}^+|\psi_\mathbf{k}^- \rangle|^2$ for a fixed azimuthal angle $\phi$ as a function of the difference of incoming ($\theta$) and outgoing ($\theta'$) polar angle.}
    \label{fig:wsm-spin1-compare}
\end{figure}

A comparison of LMC between Weyl fermions and pseudospin-1 fermions is made in Fig.~\ref{fig:wsm-spin1-compare} (a). Note that magnetoconductivity has a larger magnitude for pseudospin-1 fermions and switches its sign from positive to negative for a much smaller value of internode scattering strength. We intuitively understand this by first examining the physics at zero magnetic field. We 
compare the overlap of fermions in opposite valleys ($|\langle \psi_\mathbf{k}^+|\psi_\mathbf{k}^- \rangle|^2$), which is responsible for chirality breaking transport (chiral anomaly). This is shown in Fig.~\ref{fig:wsm-spin1-compare} (b). Clearly, pseudospin-1 fermions are more likely to get backscattered while flipping their chirality compared to Weyl fermions. The exact likelihood ($l$) is calculated to be 
\begin{align}
l=\frac{(8 + 3 \pi)}{(6 + 3 \pi)}\sim 1.12.
\end{align}
This increased likelihood of backscattering results in a quicker conductivity decrease due to internode scattering. We further analytically evaluate the zero-field conductivity using the Boltzmann formalism to 
\begin{align}
\sigma_{zz}^{s=1} = \frac{\mathrm{e}^2 v_{\mathrm{F}}^2}{ V \pi^2 \left(3 \alpha + 1\right)}, \nonumber\\
\sigma_{zz}^{s=1/2} = \frac{\mathrm{e}^2 v_{\mathrm{F}}^2}{16 V \pi^2 \left(2 \alpha + 1\right)},
\label{Eq:LMC_B0}
\end{align}
where $V= U^2\mathcal{V}/\hbar$ ($U$ being the strength of the impurity potential). The conductivity for pseudospin-1 fermions is greater in magnitude than Weyl fermions for the same set of parameters, and 
again, we note that conductivity depends more strongly on internode scattering in pseudospin-1 fermions than in Weyl fermions. In the absence of intervalley scattering, it is straightforward to evaluate that for fermions with higher pseudospin ($s>1$), the zero-field conductivity also increases with $s$. We conjecture that the conductivity in the presence of intervalley scattering ($\alpha$) should also drop more dramatically with increasing $s$ (although this must be confirmed by explicit calculations). This suggests that we may need more diagnostic tools other than relying on negative magnetoresistance studies to confirm CA in systems with higher pseudospin-$s$ fermions.

\textit{Outlook:} Condensed matter systems provide a unique platform for studying emergent fermions that otherwise have no analogs in high-energy physics. Pseudospin-1 fermions form one such example that can emerge in candidate materials with space group symmetries 
199, 214, and 220. Similarly, higher pseudospin excitations are possible as well~\cite{bradlyn2016beyond}. 
Investigating chiral anomaly and its manifestation in transport experiments can reveal fascinating properties of (pseudo)relativistic fermions beyond the standard Dirac fermions and shed light on the universality of (pseudo)relativistic fermions.  We conjecture that magnetoconductivity in the presence of intervalley scattering should quickly become negative with increasing pseudospin index, which suggests that we may need more diagnostic tools other than relying on negative magnetoresistance studies to confirm CA in systems with higher pseudospin-$s$ fermions. This will help design upcoming experiments on candidate materials where such excitations can emerge.

\bibliography{biblio.bib}

\begin{widetext}
\huge{\centering\section*{
Supplemental material to `Chiral anomaly and longitudinal magnetoconductance in pseudospin-1 fermions'}}
\normalsize

\section{Substitution and expression for system of linear equations}
\label{Appendix:Substitution and expression}
The system of linear equations of the unknown variables involved in the $\Lambda^{\chi}$ is written as follows:
\begin{multline}
\begin{bmatrix}
      a^+\\
      b^+\\
      c^+\\
      a^-\\
      b^-\\
      c^-\\
\end{bmatrix}
=
\begin{bmatrix}
\Pi^{++}T^+ & \Pi^{++}I^+ & \Pi^{++}J^+ & \Pi^{+-}I^- & \Pi^{+-}M^- & \Pi^{+-}N^-\\
\Pi^{++}I^+ & \Pi^{++}M^+ & \Pi^{++}N^+ & \Pi^{+-}T^- & \Pi^{+-}I^- & \Pi^{+-}J^-\\
\Pi^{++}J^+ & \Pi^{++}N^+ & \Pi^{++}Q^+ & \Pi^{+-}J^- & \Pi^{+-}N^- & \Pi^{+-}Q^-\\
\Pi^{-+}I^+ & \Pi^{-+}M^+ & \Pi^{-+}N^+ & \Pi^{--}T^- & \Pi^{--}I^- & \Pi^{--}J^-\\
\Pi^{-+}T^+ & \Pi^{-+}I^+ & \Pi^{-+}J^+ & \Pi^{--}I^- & \Pi^{--}M^- & \Pi^{--}N^-\\
\Pi^{-+}J^+ & \Pi^{-+}N^+ & \Pi^{-+}Q^+ & \Pi^{--}J^- & \Pi^{--}N^- & \Pi^{--}Q^-\\
\end{bmatrix}
\cdot
\begin{bmatrix}
      a^+\\
      b^+\\
      c^+\\
      a^-\\
      b^-\\
      c^-\\
\end{bmatrix}
-
\begin{bmatrix}
\Pi^{++}G^+ + \Pi^{+-}L^-\\
\Pi^{++}L^+ + \Pi^{+-}G^-\\
\Pi^{++}P^+/4 + \Pi^{+-}P^-/4\\
\Pi^{-+}L^+ + \Pi^{--}G^-\\
\Pi^{-+}G^+ + \Pi^{--}L^-\\
\Pi^{-+}P^+/4 + \Pi^{--}P^-/4\\
\end{bmatrix}
\label{Eq:System of linear equation}
\end{multline}

Substitution in the above matrix equation is written as:
\begin{align*}
\int d\theta' f^{\chi}(\theta') \cos^4(\theta'/2) &= R^{\chi}, \quad 
\int d\theta' f^{\chi}(\theta') \sin^4(\theta'/2) = S^{\chi}, \\
\int d\theta' f^{\chi}(\theta') \sin^8(\theta'/2) &= M^{\chi}, \quad 
\int d\theta' f^{\chi}(\theta') h^{\chi}(\theta') = U^{\chi}, \\
\int d\theta' f^{\chi}(\theta') \cos^8(\theta'/2) &= T^{\chi}, \quad 
\int d\theta' f^{\chi}(\theta') \sin^2(\theta') = O^{\chi}, \\
\int d\theta' f^{\chi}(\theta') h^{\chi}(\theta')\sin^2(\theta') &= P^{\chi}, \quad 
\int d\theta' f^{\chi}(\theta') h^{\chi}(\theta') \sin^4(\theta'/2) = L^{\chi}, \\
\int d\theta' f^{\chi}(\theta') h^{\chi}(\theta') \cos^4(\theta'/2) &= G^{\chi}, \quad 
\int d\theta' f^{\chi}(\theta') h^{\chi}(\theta') \cos^4(\theta'/2) \sin^2(\theta') = J^{\chi}, \\
\int d\theta' f^{\chi}(\theta') h^{\chi}(\theta') \sin^4(\theta'/2) \sin^2(\theta') &= N^{\chi}, \quad 
\int d\theta' f^{\chi}(\theta') h^{\chi}(\theta') \sin^4(\theta'/2) \cos^4(\theta'/2) = I^{\chi},\\
\int d\theta' f^{\chi}(\theta') \sin^4(\theta') &= Q^{\chi}.
\end{align*}
\section{Zero-field conductivity}\label{Sec:Zero magnetic field coductivity}
We calculate the zero-field conductivity and derive its analytical expressions for two systems, i.e., spin-1/2 and pseudospin-1 using the formalism mentioned in the main text. For $\mathbf{B} = 0 $, the the coupled Eq.~\ref{Couplled_equation} acquire the following form:
\begin{align}
\dot{\mathbf{r}}^\chi &= \frac{e}{\hbar}(\mathbf{E}\times \mathbf{\Omega}^\chi) + \mathbf{v}_\mathbf{k}^\chi \nonumber\\
\dot{\mathbf{k}}^\chi &= -\frac{e}{\hbar} ~\mathbf{E}  .
\label{Eq:Couplled_equation_B0}
\end{align}
Using this, the Boltzmann equation is simplified to:
\begin{align}
v^{\chi}_{z,\mathbf{k}} + \frac{\Lambda^{\chi}(\theta)}{\tau^{\chi}_{\mu}}=\sum_{\chi'}\int\frac{d^3\mathbf{k}'}{(2\pi)^3} \mathbf{W}^{\chi \chi'}_{\mathbf{k k'}}\Lambda^{\chi'}(\theta').
\label{Eq:MB_in_term_Wkk'_B0}
\end{align}
We solve this analytically using the ansatz defined in the main text and using the ansatz defined in Ref.~\cite{ahmad2021longitudinal,ahmad2024nonlinear} for Weyl (spin-1/2). The net scattering rate at two valleys ($\chi=\pm1$) for $B=0$ is calculated to be:
\begin{align}
\frac{1}{\tau^{\chi,s=1}_{\mu}} = \frac{4 \pi V \epsilon_{\mathrm{F}}^2 (\alpha + 1)}{3 \hbar^3 v_{\mathrm{F}}^3} \equiv \frac{8}{\tau^{\chi,s=1/2}_{\mu}}.
\label{Eq:tauinv_pm1_B0}
\end{align}
Returning back to Eq.~\ref{Eq:MB_in_term_Wkk'_B0}, it is written explicitly in terms of unknown variables involved in ansatz for spin-1 WSMs as: 
\begin{multline}
\lambda^{+} - \frac{\sin^2(\theta) \left( 10\lambda^{+} + 3a^{+} + 3b^{+} - 12c^{+} + 10\lambda^{-}\alpha + 3\alpha a^{-} + 3\alpha b^{-} - 20\alpha c^{+} + 8\alpha c^{-} \right)}{20(\alpha + 1)} \\
- \frac{\cos^4\left(\frac{\theta}{2}\right) \left( 10\lambda^{+} - 4a^{+} + b^{+} + 6c^{+} - 5v_{\mathrm{F}} + 10\lambda^{-}\alpha - 10\alpha a^{+} + \alpha a^{-} + 6\alpha b^{-} + 6\alpha c^{-} + 5\alpha v_{\mathrm{F}} \right)}{10(\alpha + 1)}\\
- \frac{\sin^4\left(\frac{\theta}{2}\right) \left( 10\lambda^{+} + a^{+} - 4b^{+} + 6c^{+} + 5v_{\mathrm{F}} + 10\lambda^{-}\alpha + 6\alpha a^{-} - 10\alpha b^{+} + \alpha b^{-} + 6\alpha c^{-} - 5\alpha v_{\mathrm{F}} \right)}{10(\alpha + 1)}=0
\end{multline}
\begin{multline}
\lambda^{-} - \frac{\sin^2(\theta) \big( 10\lambda^{-} + 3a^{-} + 3b^{-} - 12c^{-} + 10\lambda^{+}\alpha + 3\alpha a^{+} + 3\alpha b^{+} + 8\alpha c^{+} - 20\alpha c^{-} \big)}{20(\alpha + 1)} \\
- \frac{\cos^4\big(\frac{\theta}{2}\big) \big( 10\lambda^{-} - 4a^{-} + b^{-} + 6c^{-} - 5v_{\mathrm{F}} + 10\lambda^{+}\alpha 
+ \alpha a^{+} - 10\alpha a^{-} + 6\alpha b^{+} + 6\alpha c^{+} + 5\alpha v_{\mathrm{F}} \big)}{10(\alpha + 1)}\\
- \frac{\sin^4\big(\frac{\theta}{2}\big) \big( 10\lambda^{-} + a^{-} - 4b^{-} + 6c^{-} + 5v_{\mathrm{F}} + 10\lambda^{+}\alpha + 6\alpha a^{+} + \alpha b^{+} - 10\alpha b^{-} + 6\alpha c^{+} - 5\alpha v_{\mathrm{F}} \big)}{10(\alpha + 1)}=0.\\
\label{Eq:MBT_in_terms_unknown_variables_B0}
\end{multline}
Equating the coefficient of  $\cos^4\big(\frac{\theta}{2}\big),~\sin^4\big(\frac{\theta}{2}\big)$ and $\sin^2(\theta)$ in these two equations results in a system of eight coupled equations to be solved for eight unknown variables ($\lambda^{\pm}, a^\pm, b^\pm, c^\pm$). The solution obtained is as follows: $\lambda^{\pm} = 0, a^{\pm} = \frac{v_{\mathrm{F}} (\alpha-1)}{3\alpha+1}, 
b^{\pm} = a^{\pm}, c^{\pm} = 0$. 
Using Eq.~\ref{Eq:g1}, Eq.~\ref{Eq:J_formula} and Eq.~\ref{Eq:tauinv_pm1_B0} LMC is evaluated to be:
\begin{align}
\sigma_{zz}^{s=1} = \frac{\mathrm{e}^2 v_{\mathrm{F}}^2}{ V \pi^2 \left(3 \alpha + 1\right)}, \nonumber\\
\sigma_{zz}^{s=1/2} = \frac{\mathrm{e}^2 v_{\mathrm{F}}^2}{16 V \pi^2 \left(2 \alpha + 1\right)}.
\label{Eq:LMC_B0}
\end{align}
\end{widetext}

\end{document}